\documentclass[11pt]{article}

\usepackage{epsfig,graphicx}
\usepackage{latexsym}

\parskip =\medskipamount
\thicklines
\def\be{\begin{equation}}
\def\bea{\begin{eqnarray}}
\def\ee{\end{equation}}
\def\eea{\end{eqnarray}}

\newcounter{fig}

\def\d{\mbox{d}}

\def\fp{\displaystyle}
\def\ba{\begin{array}{rcl}}
\def\ea{\end{array}}

\def\o0{\overrightarrow{0}}

\begin{document}

\begin{center}
{\Large \bf Algebraic and arithmetic area for $m$ planar Brownian paths }\\[0.5cm]

{\large \bf Jean Desbois}\footnote{jean.desbois@u-psud.fr} and 
{\large \bf St\'ephane Ouvry}\footnote{stephane.ouvry@u-psud.fr}\\[0.1cm]
Universit\'e Paris-Sud, Laboratoire de Physique Th\'eorique et Mod\`eles
Statistiques\footnote{Unit\'e Mixte de Recherche CNRS-Paris Sud, UMR 8626}\\
91405 Orsay, France
\\[0.2cm]
\end{center}

\begin{abstract}
The leading and next to leading terms of the average arithmetic area $\langle S(m)\rangle$ 
enclosed by  $m\to\infty$ independent closed Brownian planar paths, with a given length $t$ and  starting 
 from and ending at the same point, is calculated. The leading term is found to be
 $\langle S(m) \rangle \sim {\pi t\over 2}\ln m$ 
 and the $0$-winding sector arithmetic area inside the $m$ paths is
  subleading in the asymptotic regime. A closed form expression 
 for the algebraic area distribution is also obtained and discussed. 
\end{abstract}

\section{Introduction}
The question of the  area, be it algebraic $A$ or arithmetic $S$, enclosed by  a closed Brownian planar path,  of a given length $t$ and starting from and ending at a given point, is an already old subject. It  started in the mid-twentieth century with the well-known Levy's law  \cite{Levy}  for the probability distribution $P(A)$ of the algebraic area. The question of the  probability distribution $P(S)$ of the arithmetic area is a much more difficult issue pertaining
 to its non local nature. Interestingly enough, the complete calculation of the first moment
 of the distribution, $\langle S\rangle=\pi t/5$,  has been only recently achieved by SLE
 technics \cite{Jeunes}. Since on the other hand the average arithmetic area
 $\langle S_n\rangle=t/(2\pi n^2)$ of the $n\ne 0$-winding sectors inside the curve has been
 known for some time   \cite{Sn} thanks to path integral technics, one can readily deduce
 from these results that the average arithmetic area $\langle S_0\rangle$ of the
 $0$-winding sectors\footnote{The path integral approach  diverges  for $n=0$
 because  it cannot distinguish zero-winding sectors inside the path from the  outside of the
 path, the latter being of infinite area and,  trivially, also a $0$-winding sector.} inside
 the curve is $\pi t/30$. A $n$-winding sector is defined  as a set of points enclosed $n$
 times by the path and a $0$-winding sector is made of points    which are either outside
 the path, or inside it  but enclosed an equal number of times clockwise and anti-clockwise.
 Note finally that,  having in mind more simple winding properties, the asymptotic probability
 distribution at large time of the angle  spanned by one path around a given point is also known
 \cite{Spitzer}.

Clearly, the random variables $S_n$  and $S_0$ are such that  $S=\sum_{n=-\infty}^{\infty}S_n$
 and $A=\sum_{n=-\infty}^{\infty}nS_n$.  They happen to be the basic objects needed to define
  quantum mechanical models where random magnetic impurities are modelised by Aharonov-Bohm
 vortices \cite{Nous}. Recently some progresses have been made on the geometrical structure
 of the $n$-winding sectors 
 thanks to numerical  
simulations of the Hausdorff dimension  of their fractal perimeter \cite{Desbois}.

The question addressed in the present work concerns the generalisation of $\langle S\rangle=\pi t
/5$ to the arithmetic area $\langle S(m)\rangle$ spanned by $m$ independent closed paths of a given length $t$  starting and returning at the same point. More precisely, one would like to have some information on the scaling of $\langle S(m)\rangle$   when $m\to\infty$.  Using again path integral technics in the line of \cite{Sn}, one will show that it is possible   to compute exactly the leading and next to leading asymptotic terms of   $\langle S(m)-S_0(m)\rangle$ where the contribution of the $0$-windings sectors inside the $m$-paths has been substracted   precisely  for the  reason discussed above in the one path case. One will  find  that the leading asymptotic term scales like $\ln m$, namely  $\langle S(m)-S_0(m)\rangle\sim {\pi t\over 2}\ln m$ with, as already said,  no information so far on $\langle S_0(m)\rangle$  inside the paths. 

To go a little bit further, one might consider a simplification of the problem  by looking at the arithmetic area of the convex envelop of the paths, a  simpler geometrical object than their actual fractal envelop. It is  known  \cite{One} that the area of the convex envelop of one path is  $\langle S\rangle_{\rm convex}= \pi t/2$. Recently, the asymptotic behavior for $m$ paths when $m\to\infty$ has been found \cite{Julien} to be precisely $\langle S(m)\rangle_{\rm convex}\sim  {\pi t\over 2}\ln m$. In light of the identical scaling obtained for $\langle S(m)-S_0(m)\rangle$, it  means that, in geometrical terms, the $m$ paths tend to  fully  occupy the area available inside their convex envelop when $m$ is large. Clearly, $\langle S(m)\rangle_{\rm convex}$ being by construction bigger than the actual $\langle S(m)\rangle$, the latter  lies between $\langle S(m)-S_0(m)\rangle$ and $\langle S(m)\rangle_{\rm convex}$. Now, since the leading asymptotic behavior of both the lower and upper bounds are found to be equal,  in the asymptotic regime one concludes  necessarily that $\langle S(m)\rangle\sim  {\pi t\over 2}\ln m $  and that  $\langle S_0(m)\rangle$ is  subleading\footnote{This can be easily understood by noticing that  $0$-winding sectors tend to disappear when more and more paths overlap as $m$ increases, which also implies that  $n$-winding indices  tend to increase.}.

\section{Algebraic area distribution}

As a warm up, let us consider the  generalization of Levy's law   to  $m$ independent paths.  
Let us first start with one   path of length $t$, starting from and ending to a given point $\vec r$, so that  $\vec r (0) = \vec r (t)=\vec r$. The algebraic area enclosed by this path is $A=\vec k .\int_0^t { \vec r(\tau )\wedge {d \vec r(\tau )}/ 2} $ where $\vec k$ is the unit vector perpendicular to the plane. The path integral
 leads to
\be
 \langle e^{i B A  } \rangle  = \frac{{\cal G}_B (\vec r , \vec r)}
   {{\cal G}_0 (\vec r , \vec r )}   \label{a1} \ee 
 where ${\cal G}_0 (\vec r , \vec r)      =   1/(2\pi t)  $  and
 \be  {\cal G}_B (\vec r , \vec r)   \equiv 
 \int_{\vec r (0) = \vec r}^{\vec r (t) = \vec r}{\cal D}\vec r (\tau) {\displaystyle e^{-\frac{1}{2}\int_0^t 
   \dot{\vec r} ^2 (\tau) \d \tau  + i B \vec k  .
 \int_0^t  {\vec r(\tau )\wedge {d \vec r(\tau )}\over 2} } } 
         = \frac{1}{2\pi t} \frac{ \frac{B t}{2}  }{\sinh\left( \frac{B t}{2}\right) }   
\ee

In (\ref{a1})  the average $\langle \quad \rangle    $ has been made over the set $C$ of all paths of length $t$  starting from and ending at  $\vec r$. As a result
${\cal G}_B$ is  the Landau propagator of a charged particle in an uniform
 magnetic field $\vec B=B\vec k$.
  Fourier transforming, the probability distribution of $A$ -the Levy's law- is
 
\be\label{p1A} 
P(A) =\frac{1}{2 \pi} \int_{-\infty}^{+\infty}  \frac{ \frac{B t}{2}  }
 {\sinh\left( \frac{B t}{2}\right) } e^{- i B A} \d B =
 \frac{\pi }{2 t}\frac{1}{ \cosh^2 \left( \frac{\pi A}{t}   \right)} 
\ee

In the case of $m$  paths, 
 one should compute instead of (\ref{p1A})
\be\label{pmA} 
P_m(A) =\frac{1}{2 \pi} \int_{-\infty}^{+\infty} 
 \left( \frac{ \frac{B t}{2}  }
 {\sinh\left( \frac{B t}{2}\right) } \right)^m
 e^{- i B A} \d B 
\ee

 The integral in (\ref{pmA}) can be performed  by rewriting
 ${1/ \sinh(Bt/2)^m}=2^m{e^{-mBt/2}/ (1-e^{-Bt})^m}$ and by expanding, when the integration
 variable $B>0$,  the denominator in powers of  $e^{-Bt}$. One obtains finally
\be\label{dir} 
P_m(A)={m!\over 2\pi t}\sum_{k=0}^{\infty}({k+m-1\atop k})
({1\over(k+m/2+i A/t)^{m+1}}+{\rm cc})
\ee 
  where the complex conjugate term corresponds to the $B<0$ integration (it  amounts to set
 $A\to -A$ in the  $B>0$ integration result).
  
Noticing that $A$ is the sum of $m$ independent random variables each satisfying  Levy's
 law,  one expects $A$ to be gaussian when $m$ becomes large. Indeed, one can observe that, when $m \to \infty $, the main contribution  to
 the integral in (\ref{pmA}) comes from small $B$ values. Thus,   
 $$\fp \left( \frac{ \frac{B t}{2}  }
 {\sinh\left( \frac{B t}{2}\right) } \right)^m \sim \left(   
  \frac{1}{ 1+ \frac{B^2t^2}{24} } \right)^m \sim e^{-m  \frac{B^2t^2}{24}  }.   $$
  Rescaling the   area  as $\fp A'={A}/({t \sqrt{m}})$, (\ref{pmA}) leads to
\be\label{pinfA'}
P_\infty (A') = \sqrt{\frac{6}{\pi }} e^{-6A'^2}
\ee
 In terms of  $A'$, we get from (\ref{dir}) or from contour integration, 
\be P_2(A')=  \frac{\pi \sqrt{2}}{\sinh^2(\pi\sqrt{2} A')}\left(\pi\sqrt{2} A'
\coth (\pi\sqrt{2} A')  -1     \right)  
     \label{p2A'} \ee
\be P_3(A')=   \frac{\pi\sqrt{3}}{2 \cosh^2 (\pi\sqrt{3}A')}\left( 3-6\pi\sqrt{3} A' \tanh (\pi\sqrt{3} A') -
 ((\pi\sqrt{3} A')^2 + \frac{\pi^2}{4}  )
 (1-3 \tanh^2 (\pi\sqrt{3} A'))
\right)                      \label{p3A'} 
\ee

\begin{figure}
\begin{center}
\includegraphics[scale=.40,angle=0]{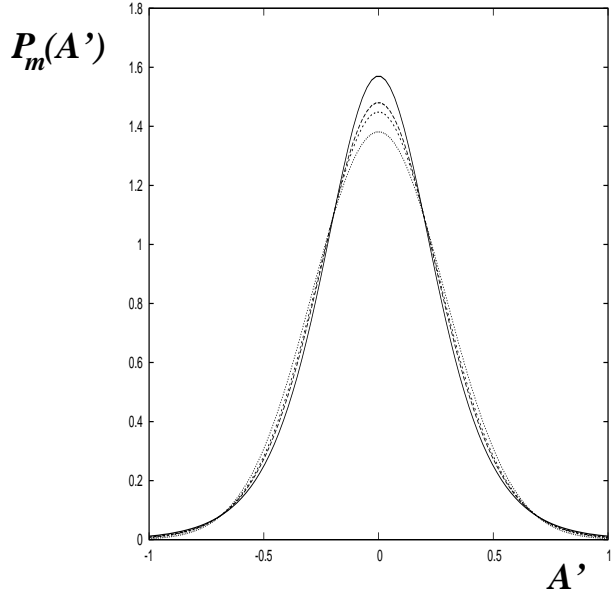}
\caption{The distribution $P_m(A')$ for the rescaled algebraic area $A'=A/(t\sqrt{m})$. 
 Starting from the curve with the greatest maximum $m=1$ {(Levy's law  eq.(\ref{p1A}))}, $m=2$ (eq. (\ref{p2A'})), $m=3$ (eq. (\ref{p3A'})), $m=\infty$ (eq. (\ref{pinfA'})). }
\label{f1}
\end{center}
\end{figure}

Figure \ref{f1} clearly shows that  $P_m(A')$ converges quickly to the gaussian 
 $ P_\infty (A')$ when $m$ becomes large. In particular $A$ scales like $\sqrt{m}$ when 
   $m \to \infty $. 
   The scaling will be different for the arithmetic area discussed in the next section.

 \section{Winding properties and arithmetic area for $m$ Brownian paths}\label{ari}
 
\subsection{The case of  one path: notations and known results.}\label{ari1}

    The arithmetic area enclosed by a planar Brownian path is closely related to its winding
  properties. Let us again consider  a path of length $t$, starting from and ending at  $\vec r$,  and let us set $\theta $ to be the angle wounded by the path
   around a fixed point, say the origin $O$. Again the average $\langle e^{i \alpha \theta  }\rangle$ is made over $C$ 
\be
 \langle e^{i \alpha \theta  } \rangle  = \frac{{\cal G}_\alpha (\vec r , \vec r)}
   {{\cal G}_0 (\vec r , \vec r)}   \label{b1} \ee
 where   
 \be {\cal G}_\alpha (\vec r , \vec r)   \equiv 
 \int_{\vec r (0) = \vec r}^{\vec r (t) = \vec r}
 {\cal D}\vec r (\tau)  e^{-\frac{1}{2}\int_0^t 
   \dot{\vec r} ^2 (\tau) \d \tau  +i \alpha  \int_0^t  \dot\theta (\tau) \d \tau  } 
         = \frac{1}{2\pi t} e^{-\frac{r^2}{t}} \sum_{k=-\infty}^{+\infty}
  I_{\vert k- \alpha \vert } \left(  \frac{r^2}{t}  \right) \ee
  By symmetry the average depends only on $r$ with
${\cal G}_\alpha$ being  the propagator of a charged particle coupled to a vortex
 at the origin  (the  $I_{\vert k- \alpha \vert }$'s are modified Bessel functions). Obvious symmetry and periodicity considerations
 such as  ${\cal G}_\alpha  = {\cal G}_{\alpha +1}  = {\cal G}_{1- \alpha } $ allow to restrict to  $0 \le  \alpha \le 1$. We also  set   $\fp {r^2}/{t}\equiv x $
  (so that $2\pi r \d r =\pi t \d x$)
  and  $\langle e^{i \alpha \theta  } \rangle  \equiv G_\alpha (x) $ with
  \be\label{gax}
   G_\alpha (x) =e^{-x} \sum_{k=-\infty }^{+\infty } I_{\vert  k- \alpha  \vert} (x)
\ee
so that $G_0(x)=1$.
  
To integrate over $\vec r $ -while keeping the position of the vortex   fixed at the origin- is the same as to
 integrate over the    vortex position -while keeping the     starting and ending
  point of the path fixed. Therefore it amounts to count the arithmetic areas $S_n$ of the
  $n$-winding sectors, the 0-winding sector included
\be\label{e1}
 \int_0^\infty \pi t \d x G_\alpha (x) = \sum_{n \ne 0} 
 \langle S_n  \rangle e^{i \alpha 2 \pi n} +  \langle \tilde{S}_0  \rangle 
 \ee
This integral  diverges since, as already stressed in the introduction,  $\tilde{S}_0$ is the sum of $S_0$, the arithmetic area  of  the 0-winding sectors enclosed by the path, and of the area of the outside of the path, which is infinite. However, since formally for $\alpha =0$, 
\be\label{e2}
  \int_0^\infty \pi t \d x G_0 (x) = \sum_{n \ne 0} 
 \langle S_n  \rangle  +  \langle \tilde{S}_0  \rangle 
\ee 
then
\be
Z_{\alpha } \equiv \pi t \int_0^\infty \d x \left(1- G_\alpha (x)\right) = \sum_{n \ne 0} 
 \langle S_n  \rangle \left(1- e^{i \alpha 2 \pi n}\right) \label{Z1} \ee
 is finite. One deduces 
 
 \be \langle S_n  \rangle   = - \int_0^1 Z_{\alpha } e^{-i \alpha 2 \pi n}  \d \alpha        
   \quad  \quad n\ne 0  \ee
   and 
\be  \langle S - S_0  \rangle       \equiv\sum_{n \ne 0} 
 \langle S_n \rangle = \int_0^1 Z_{\alpha} \d \alpha \label{aire}   
\ee

Using the Laplace transform of the modified Bessel functions, we readily recover
\be
 Z_{\alpha }  = \pi t \alpha (1 - \alpha )  \ee
 \be
  \langle S_n  \rangle   =    \frac{t}{2 \pi n^2}             \quad  \quad n\ne 0  
 \ee
 \be
  \langle S - S_0  \rangle      = \frac{\pi t}{6} 
\ee

\subsection{The  case of $m$ independent paths. }\label{ari2}

It is easy to
 realize that $  {\cal G}_\alpha (\vec r , \vec r) ^m$   provides the appropriate
 measure for counting the sets of $m$ closed paths of length $t$ starting from and ending at $\vec r$. 
Following the same line of reasoning  as in section \ref{ari1}, we get
\be
Z_{\alpha }(m) \equiv \pi t \int_0^\infty \d x \left(1-  G_\alpha (x)^m \right)
 = \sum_{n \ne 0} 
 \langle S_n(m)  \rangle \left(1- e^{i \alpha 2 \pi n} \right) \label{Zm} \ee
 and
 \be
 \langle S(m) - S_0(m)  \rangle       \equiv \sum_{n \ne 0} 
 \langle S_n(m) \rangle = \int_0^1 Z_{\alpha}(m) \d \alpha   \label{airem}
\ee
where $S(m)$, $S_n(m)$  and $S_0(m)$ stand respectively for the total, $n$-winding sectors   and $0$-winding sectors arithmetic areas enclosed by the $m$ paths. { For example a sector of points which have been enclosed once by one path in the clockwise direction, twice by another path in the anticlockwise direction, and not enclosed by the other $m-2$ paths, has winding number $n=-1+2=1$}.

To find the  leading behavior of $S(m)$ in the large $m$ limit, one  has   to evaluate $ G_\alpha(x) ^m$ when $m \to \infty$, and so one needs  a tractable
 expression for  $G_\alpha (x)$. Starting from 
(\ref{gax})
and using again the Laplace transform of  Bessel functions, we get
\be\label{dga}
\frac{\d G_\alpha (x)}{\d x}=\frac{2}{\sqrt{\pi}} \sin (\pi \alpha) e^{-2x} (2x)^{\alpha -1}
  U(\alpha -\frac{1}{2}; 2 \alpha ; 2x)
\ee
where $U$ is the degenerate hypergeometric function \cite{abra}
\be\label{hgy} 
U(a;b;z)= \frac{z^{-a}}{\Gamma (a)} \int_0^\infty e^{-t} t^{a-1} 
\left(1+ \frac{t}{z} \right)^{b-a-1} \d t
\ee
It is easy to verify that $G_\alpha (x)=G_{1 - \alpha } (x)$, as it should.
 
From (\ref{dga})
\be\label{inf}
 G_\alpha (\infty )=G_\alpha (0 )+\int_0^\infty \frac{\d G_\alpha (x)}{\d x} \d x =1
\ee
as it should far away from the origin (one has used that  when $\alpha \ne 0 $, $G_\alpha (0)=0$).
More precisely, since for $z$ large $U(a;b;z) \sim z^{-a} $, one has
\be\label{ginf}
1- G_\alpha (x ) \sim  \frac{\sin ( \pi \alpha )}{ \sqrt{2 \pi x}} e^{-2x}
 \quad \mbox{ when } \quad x \to \infty
\ee
On the other hand, for $\fp  0 < \alpha < {1}/{2}$,
\be\label{gto0}
 G_\alpha (x ) \sim \frac{\left( \frac{x}{2} \right)^\alpha}{\Gamma ( \alpha + 1) } 
 \quad \mbox{ when } \quad x \to 0
\ee

Changing variable $\fp 2 x = {y \ln m}$ in (\ref{Zm}) 
 \be\label{Zma}
Z_{ \alpha}(m)=\frac{\pi t}{2} \ln m \int_0^\infty \d y
\left( 1 -\left(G_\alpha (\frac{y \ln m}{2})\right) ^m  \right)
\ee
one sees that  in the limit $m \to \infty$ 
\begin{itemize}
\item when $y=0$, that is ${y \ln m}=0$,  $\left(G_\alpha  \right)^m =0$

\item when $y>0$, that is $y \ln m \to \infty$,  the asymptotics 
 (\ref{ginf}) can be used so that
\be\label{toto}
   \left( G_\alpha (\frac{y \ln m}{2}) \right)^m  \to   \left(   
 1 -  \frac{ \sin ( \pi \alpha )}{ \sqrt{ \pi y \ln m} } m^{-y} 
\right)^m 
 \sim  {\displaystyle e}^{\displaystyle - \frac{ \sin ( \pi \alpha )}{ \sqrt{ \pi y \ln m} } m^{1-y} }   
\ee 
\end{itemize}
It follows  that
\be \left( G_\alpha (\frac{y \ln m}{2}) \right)^m \to_{m\to\infty}  \Theta (y-1)\ee 
where
 $\Theta$ is the Heaviside function.
Finally
\be
 Z_{\alpha}(m)   \sim \frac{\pi t }{2} \ln m  \ee
 and
 \be \langle S(m) -S_0 (m)     \rangle    \sim  \frac{\pi t }{2} \ln m  
  \ee
As already discussed in the Introduction, it means that  since
  $\langle S(m) \rangle_{\rm{convex}}\sim  {\pi t\over 2 } \ln m $,  at leading order 
\be \langle S(m)      \rangle    \sim  \frac{\pi t }{2} \ln m  
  \ee  
  and, necessarily, 
  $ \langle S_0 (m)  \rangle  $ is subleading.

\begin{figure}
\begin{center}
\includegraphics[scale=.40,angle=0]{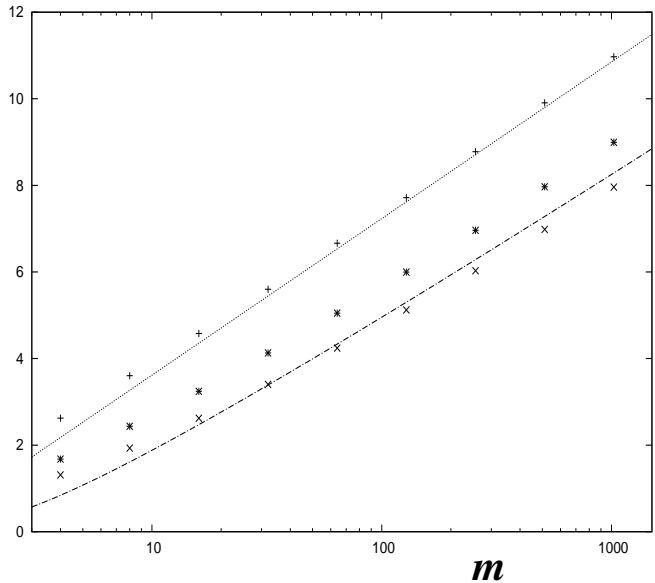}
\caption{The average arithmetic area: 
i) 
numerical simulations (10000 events) of closed random walks of $10^6$ steps
 on a square lattice; $m = 4, 8, 16, ..., 1024$; \ \
 +:   $  \langle S(m) \rangle_{\rm{convex}}/t$, \ \ 
  *:   $  \langle S(m) \rangle/t $, \ \ 
  $\times $:  $  \langle S(m) - S_0(m) \rangle/t $ \ 
ii) 
   analytical results  for  $ \langle S(m) - S_0(m) \rangle/t $;  upper continuous line: 
${\pi  } \ln (m)/2 $ (leading order);  \ lower dotted line: eq. (\ref{result}) that includes the subleading
 corrections.
}\label{f2}
\end{center}
\end{figure}

In Figure \ref{f2}, numerical simulations (random walks on a square lattice)  for
 $ \langle S(m) \rangle_{\rm{convex}}/t$,
 $\langle S(m) \rangle/t $ and  
 $\langle S(m) -  S_0 (m) \rangle /t$ are displayed. 
 One sees that if $ \langle S(m) \rangle_{\rm{convex}}/t$ converges rapidly to   
  ${ \pi\over 2 } \ln m $, this is not the case for 
 $ \langle S(m) \rangle/t $ and
 $ \langle S(m) -  S_0 (m) \rangle/t $. It means that subleading corrections are 
 needed. They originate  from the fact that, when ${m}$ is large but not  infinite,  $( G_\alpha)^m$ deviates from a Heaviside function. One should compute
\be\label{comp}
{1\over t}\langle S(m) -  S_0 (m) \rangle =  \frac{ \pi }{2} \ln m + \frac{ \pi }{2} \ln m\left(
 c_1+c_2
\right)
\ee
with
\be
     c_1  = \int_0^1 \d \alpha  \int_1^\infty \d y 
  \left( 1-   \left(  G_\alpha (\frac{y \ln m}{2}) \right)^m  \right)  \ee
 \be   c_2    = - \int_0^1 \d \alpha  \int_0^1  \d y
  \left(  G_\alpha (\frac{y \ln m}{2}) \right)^m  \ee
For $y>1$,
 \be 1-  \left(  G_\alpha (\frac{y \ln m}{2}) \right)^m \sim m \frac{\sin (\pi \alpha)}
{\sqrt{\pi y \ln m}} m^{-y} \ee  so that
\be\label{c1}
\frac{ \pi }{2} \ln m \; c_1 \sim \frac{1}{\sqrt{ \pi \ln m}}
\ee
For $0< y\le 1$, 
 \be \left(  G_\alpha (\frac{y \ln m}{2}) \right)^m \sim e^{-a m^{1-y}} \ee
 with $a ={\sin (\pi \alpha)}/{\sqrt{\pi y \ln m}}\approx 
{\sin (\pi \alpha)}/{\sqrt{\pi  \ln m}}$ since when $m$ is large, $y$ is peaked to $1$.
Changing  variable to \ $\fp a \; m^{1-y } = z $ \  and integrating  over $z$, then over 
 $\alpha $, leads to
\be\label{c2}
\frac{ \pi }{2} \ln m \; c_2 \sim -\frac{\pi }{4} \ln \ln m - 
\frac{ \pi }{2} \left( \ln \sqrt{4 \pi } -C  \right)
 - \frac{1}{\sqrt{ \pi \ln (m)}} +
 {\rm subleading}
\ee
where $C$ is the Euler constant.
Collecting all  terms, we finally obtain
\be\label{result}
 \frac{1}{t} \langle S(m) -  S_0 (m) \rangle =  \frac{ \pi }{2} \ln m 
  -\frac{\pi }{4} \ln \ln m
 - \frac{ \pi }{2} \left( \ln \sqrt{4 \pi } -C  \right) + {\rm subleading}
\ee
 One sees in Figure \ref{f2} that the subleading corrections  greatly
 improve the fit: the lower dotted line (\ref{result}) is indeed not  far from the numerical
 data (the agreement is of course not entirely perfect but further subleading corrections seem hard
 to reach).

\section{Conclusion}

 In conclusion one has established that the leading behavior of the average arithmetic area $\langle S(m)\rangle$ 
enclosed by  $m\to\infty$ independent closed Brownian planar paths is ${\pi t\over 2}\ln m$.  The  algebraic area, on the other hand,  scales like $t\sqrt{m}$. One should stress that the quite different asymptotic behaviors  pertain to the essentially different nature of the areas considered: the algebraic area is additive, which is not the case of the arithmetic area. On another front, it remains a real challenge  to get some information on the subleading asymptotic behavior of the  $0$-winding sector area $\langle S_0(m)\rangle$. Again path integral technics  are not adapted to this case, whereas SLE machinery should  in principle work. 



\begin{thebibliography}{99}


\bibitem{Levy} P. L\'evy,  Processus Stochastiques et Mouvement Brownien, Paris, Gauthier-Villars (1965); in Proceedings Second Berkeley Symposium on Mathematical Statistics and Probability, University of California Press (1951) 171 

\bibitem{Jeunes} C. Garban and J. A. Trujillo Ferreras, Commun. Math. Phys. 264 (2006) 797

\bibitem{Sn} A. Comtet, J. Desbois and S. Ouvry, J. Phys. {\bf A 23} (1990)
3563

 \bibitem{Thesis} W. Werner, Th\`ese, Universit\'e Paris 7 (1993) and Probability
Theory and Related Fields (1994) 111

\bibitem{Spitzer} F. Spitzer, Trans. Amer. Math. Soc., 87 (1958)  187-197

\bibitem{Nous} J. Desbois, C. Furtlehner and S. Ouvry,
Random Magnetic Impurities and the Landau Problem,
Nuclear Physics B[FS] {\bf 453} (1995) 759


\bibitem{Desbois} J. Desbois and S. Ouvry,  JSTAT (2008) P08004

\bibitem{One} M. El Bachir, Th\`ese, Universit\'e Paul Sabatier, Toulouse (1983)


\bibitem{Julien} J. Randon-Furling,  S. Majumdar and A. Comtet,  Phys. Rev. Lett.,
v-103 (2009) 140602; see also by the same authors J. Stat. Phys. 138 (2010) 955

\bibitem{abra} M. Abramowitz and I. Stegun, Handbook of mathematical functions, New york,
 Dover Publications (1965)



\end{thebibliography}
\end{document}